\documentclass[journal]{vgtc}                

\usepackage{mathptmx}
\usepackage{graphicx}
\usepackage{times}
\usepackage[lined,algonl,boxed]{algorithm2e}
\usepackage{amssymb,amsmath}
\usepackage{multirow}
\usepackage{subfigure}
\usepackage{color}
\usepackage{paralist}
\usepackage{tikz}
\usepackage{multirow}
\usepackage{color}
\usepackage{ulem}
\usepackage{setspace}
\usepackage{microtype,hyphenat,balance}
\usepackage{stfloats}
\usepackage{array}
\usepackage{url}
\usepackage{bm}
\newcommand{\shimei}[1]{\textcolor{black}{#1}}
\newcommand{\kgsedit}[1]{\textcolor{black}{#1}}

\newcommand{\revision}[1]{\textcolor{black}{#1}}
\newcommand{\doc}[1]{\textcolor{black}{#1}}
\onlineid{226}

\vgtccategory{Research}

\vgtcinsertpkg

\title{An Uncertainty-Aware Approach for Exploratory Microblog Retrieval}

\author{Mengchen Liu, Shixia Liu, Xizhou Zhu, Qinying Liao, Furu Wei, and Shimei Pan}

 \authorfooter{
\item
M. Liu and S. Liu are with Tsinghua University. E-mail: simon900314 @139.com, shixia@tsinghua.edu.cn. S. Liu is the corresponding author.

\item
X. Zhu is with USTC. E-mail: ezra0408@mail.ustc.edu.cn.
\item

Q. Liao and F. Wei are with Microsoft. E-mail: \{qiliao,fuwei\}@microsoft.com.

\item
S. Pan is with University of Maryland, Baltimore County. E-mail: shimei@umbc.edu
 }

\abstract{
Although there has been \doc{a great deal of} interest in analyzing customer opinions and breaking news in microblogs, progress has been hampered by the lack of an effective mechanism to discover and retrieve data of interest from microblogs.
To address this problem, we have developed an uncertainty-aware visual analytics approach to retrieve salient posts, users, and hashtags.
We extend an existing ranking technique to compute a multifaceted retrieval result: the mutual reinforcement rank of a graph node, the uncertainty of each rank, and the propagation of uncertainty among different graph nodes.
To illustrate the three facets, we \doc{have} also design\doc{ed} a composite visualization with three visual components: a graph visualization, an uncertainty glyph, and a flow map.
The graph visualization with glyphs, the flow map, and the uncertainty analysis together enable analysts to effectively find the most uncertain results and interactively refine them.
We have applied our approach to several Twitter datasets. Qualitative evaluation and two real-world case studies demonstrate the promise of our approach for retrieving high-quality microblog data.\looseness=-1

}

\keywords{microblog data, mutual reinforcement model, uncertainty modeling, uncertainty visualization, uncertainty propagation.} 

\teaser{
\centering
\centering
  \includegraphics[width=\linewidth]{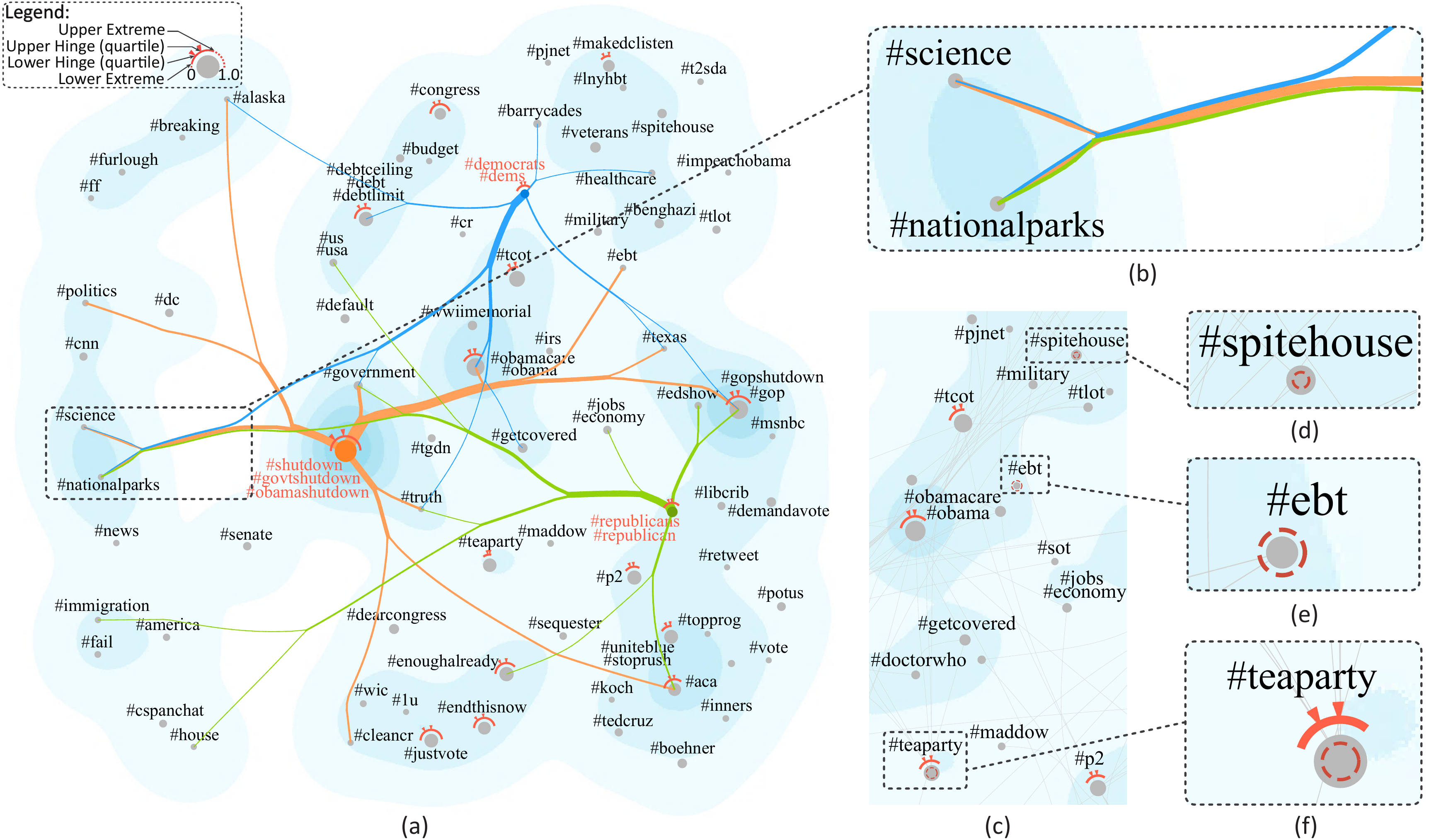}
    \vspace{-7mm}
  \caption{
  Exploratory retrieval of the government shutdown dataset: (a) \revision{the hashtag graph} with uncertainty and its propagation; (b) uncertainty propagation; (c)-(f) interactive ranking refinement results.
  }
\vspace{-1mm}
\label{fig:shutdownoverview}
}

\begin{document}


{
\fontsize{8}{8} 

\firstsection{Introduction}
\maketitle
Microblogs such as Twitter and Facebook are among the most popular platforms \doc{for} people \doc{to} share their daily observations and thoughts, including personal status updates and opinions regarding products or government policies.
Since the crowd in microblogs provides many individual comments/opinions that were not available before,
businesses and organizations have begun to leverage microblogs to profile customers, derive brand perception, gauge citizen sentiments, and predict the stock market~\cite{Liu2012_survey,Ruiz2012_WSDM,tang2013portraying,Zhao2014_KDD}.
For example, retailers track and examine relevant microblog posts to understand customer opinion toward their products and services.
In spite of the growing interest in quickly analyzing customer opinions or breaking news in \doc{microblogs,} progress has been hampered by the lack of an effective mechanism to retrieve data of interest from microblogs.

For this reason, researchers have developed a number of microblog retrieval methods~\cite{Bosch2013_TVCG,Efron2011_survey}.
The main goal is to generate a list of \emph{k} microblog posts that are relevant to the information \doc{needs} represented by a query \emph{q}.
Although these methods have successfully retrieved relevant posts, they have two major drawbacks.
First, the unique characteristics of microblog data are not comprehensively considered \revision{to improve retrieval performance.}
Posts, users, and hashtags are three key dimensions of microblog data.
These dimensions are not independent, as they often influence one another.
For example, a post published by an influential user and labeled with a popular hashtag tends to be salient.
However, the existing approaches to microblog retrieval do not tightly integrate the three dimensions and do not take advantage of the relationships among them.
These approaches either consider only the \shimei{post or} treat it as the primary dimension and \doc{the} others as secondary dimensions to filter posts.
For example, ScatterBlogs2~\cite{Bosch2013_TVCG} finds posts of interest by checking whether the posts contain a certain hashtag.
Second, existing methods do not address uncertainty in  the retrieval models.
\shimei{Improving} the modeling and presentation of uncertainty can help describe retrieved data \shimei{more} accurately, which can in turn assist analysts in making more informed decisions\shimei{~\cite{Correa2009_VAST,Sun2013survey,Wu2012_InfoVis}}.\looseness=-1

To address the above problems, we have developed an uncertainty-aware microblog retrieval toolkit, MutualRanker, \revision{to estimate uncertainty introduced by the analysis algorithm as well as to quickly retrieve salient posts, users, and hashtags}.
Since posts are \shimei{shared} and propagated on social networks, the authority of an author and the popularity of a hashtag play an important role in determining the importance of a post and vice versa.
Accordingly, we formulate uncertainty-aware microblog retrieval using an uncertainty-based mutual reinforcement graph model (MRG)~\cite{Duan2012_coling,Wei2008_SIGIR}, where the content quality of posts, the social influence of users, and the popularity of hashtags mutually reinforce one another.
We adopt a Monte Carlo sampling method to solve MRG because of its effective local update mechanism, fast convergence, and probability-based uncertainty \shimei{formalization}~\cite{Avrachenkov2007}.
\shimei{A} dispersion-based measure is used to estimate the uncertainty generated by the Monte Carlo sampling method.
In addition, we model the uncertainty propagation as a Markov chain.\looseness=-1

To help analysts understand \doc{the} retrieved data, we have designed a composite visualization~\cite{Javed2012_PacificVis}.
Specifically, a density-based graph visualization \doc{has been} developed to visually illustrate posts, users, hashtags, and their relationships.
An uncertainty glyph and a flow map are employed to represent uncertainty and its propagation on a graph (Fig.~\ref{fig:shutdownoverview}).
The three visualization components, together with the uncertainty analysis\doc{,} enable analysts to quickly detect the most uncertain results and interactively resolve them.
\doc{The} Monte Carlo sampling method is \doc{then} used to incrementally modify the ranking results to meet \doc{user} needs.

In summary, our work presents three technical contributions:
\begin{compactitem}
\item \textbf{\normalsize An uncertain-aware microblog retrieval model} that extracts salient posts, users, and hashtags.
This model also computes the associated uncertainty and its propagation among graph nodes.

\item \textbf{\normalsize A composite visualization} that enables users to understand the three-level, mutual reinforcement ranking results, the associated uncertainty, and uncertainty propagation patterns.\looseness=-1

\item \textbf{\normalsize A visual analytics system} that helps users quickly retrieve data of interest, as well as analyze and understand the ranking results in an interactive and iterative process.
\end{compactitem} 

\section{Related Work}\label{sec:related-work}

\subsection{Microblog Retrieval}
In the field of data mining, a number of approaches have been proposed to retrieve data from microblogs.
A comprehensive survey was presented by Cherichi and Faiz~\cite{Cherichi2013relevant}.
\doc{Most recent work can be} categorized into two groups: vector-space-based approaches and link analysis approaches.

The vector-space-based approach employs two feature vectors to represent a query and a post.
\doc{A} similarity measure (e.g., cosine similarity) is \doc{then} adopted to estimate the similarity between the post and the query.
There \doc{have been} some recent research efforts \doc{that exploit} additional structural features such as URLs and hashtags to enhance retrieval performance~\cite{Alhadi2011livetweet,Luo2012_AAAI,McCreadie2013_OAIR}.


\revision{Recently, \shimei{to take advantage of the link structure of social networks,} researchers have introduced the PageRank algorithm~\cite{Brin1998anatomy} \shimei{in} microblog retrieval.}
For example, TwitterRank~\cite{Weng2010_WSDM} adopts the follower-followee link structure and the PageRank algorithm to identify influential users.
Duan et al.~\cite{Duan2012_coling} modeled the \doc{tweet-ranking} problem as \doc{an} MRG~\cite{Wei2008_SIGIR}, where the social influence of users and the content quality of tweets mutually \doc{reinforce} each other.
Specifically, the post graph, the user graph, and the hashtag graph\doc{,} as well as the relationships between the three graphs\doc{,} were used to retrieve salient posts, users\doc{,} and hashtags.
We extend this approach by explicitly modeling the uncertainty of the ranking result, as well as its propagation on the tweet/user/hashtag graph.\looseness=-1
%

In the field of visual analytics, \doc{a great deal of} research has been conducted \doc{on} visually \doc{analyzing} microblog data.
The methods applied include event detection~\cite{Marcus2011_CHI}, topic extraction and analysis~\cite{Liu2014_VAST,Sun2014_TVCG,Xu2013_TVCG}, information diffusion~\cite{Cao2012_InfoVis,Zhao2014_TVCG}, sentiment analysis~\cite{Wu2014_TVCG,Wu2010_Infovis}, and revenue/stock prediction~\cite{Lu2014_CGA,Ruiz2012_WSDM}.
However, few studies have focused on microblog retrieval.\looseness=-1

Bosch et al.~\cite{Bosch2013_TVCG} developed ScatterBlogs2 to extract microblog posts of interest.
\shimei{It} allows analysts to build \shimei{customized} post filters and classifiers interactively.
\doc{These} filters and classifiers are \doc{then} utilized to support real-time post monitoring.
\shimei{In post filtering,} the post dimension \shimei{is considered} the primary dimension and the \shimei{hashtag the} secondary \shimei{dimension.}
\doc{In contrast}, we tightly integrate the posts, users, and hashtags in the MRG model and use the model to retrieve high-quality microblog data.
Moreover, \shimei{we also model} uncertainty in the retrieval process.
\revision{\shimei{Since analysts can interactively refine the model, we can further} improve retrieval quality by leveraging the uncertainty formalization and analysts' knowledge.}

\subsection{Interactive Uncertainty Analytics}
\shimei{Frequently, uncertainty is} introduced into visual analytics when data is acquired, transformed, or visualized~\cite{Correa2009_VAST,Liu2014survey,Lodha1996_visualization}.
A number of uncertainty analysis methods have been proposed, which can be categorized into two groups: uncertainty visualization and uncertainty modeling.\looseness=-1

Many studies on uncertainty visualization have been conducted in the field of geographic visualization and scientific visualization~\cite{Pang1997_TVCJ,Skeels2010revealing,Thomson2005}.
Typical uncertainty representation techniques include the addition of glyphs and geometry, the modification of geometry and attributes, animation, sonification, and psycho-visual approaches~\cite{Pang1997_TVCJ}.
Recently, researchers are increasingly interested in the design of uncertainty representations for information visualization and visual analytics.
For example, Collins et al.~\cite{Collins2007_VUL} designed two alternatives, the gradient border and the bubble border, to illustrate uncertainty in lattice graphs.
Wu et al.~\cite{Wu2010_Infovis} developed a circular wheel representation and subjective logic to convey uncertainty in customer review analysis.
Slingsby et al.~\cite{Slingsby2011_InfoVis} utilized bar charts to reveal the uncertainty associated with geodemographic classifiers.
To represent uncertainty in aggregated vertex sets, Vehlow et al.~\cite{Vehlow2013_InfoVis} considered the lightness \doc{and shape} of the node.
Chen et al.~\cite{Chen2015_TVCG} adopted the uncertainty histogram to explore uncertainty in the context of a multidimensional ensemble dataset.
Compared with these methods, MutualRanker not only visualizes uncertainty, but also its propagation on a graph.
We also \shimei{support users to interactively} modify the uncertain result.

Another type of uncertainty visualization \doc{represents} the uncertainty in the analysis process.
Zuk and Carpendale~\cite{Zuk2007_SG} studied issues related to uncertainty in reasoning and \shimei{determined} the type of visual support required.
Correa et al.~\cite{Correa2009_VAST} developed a framework to represent and quantify the uncertainty in the visual analytics process.
Wu et al.~\cite{Wu2012_InfoVis} extended this framework to show the uncertainty flow in the analysis process.
By contrast,
our work aims to model uncertainty in microblog retrieval.
We focus on visually illustrating topological \shimei{uncertainty} propagation on a graph and on designing an iterative visual analytics process to actively engage analysts in reducing overall uncertainty.

\begin{figure*}[t]
  \centering
  \includegraphics[width=\linewidth]{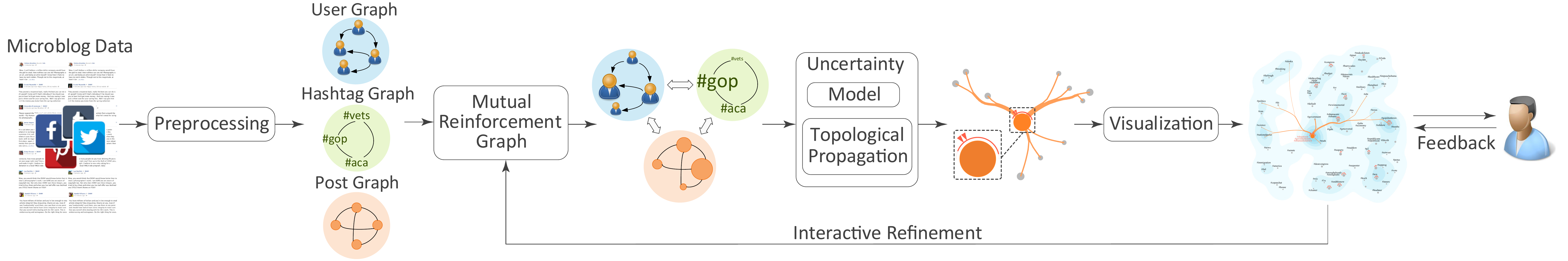}
  \centering
  \caption{MutualRanker pipeline.}
  \label{fig:overview}
  \vspace{-5mm}
\end{figure*}

Probability theory, fuzzy set theory, rough set theory, and evidence theory are four major approaches to model uncertainty~\cite{Zimmermann2001fuzzy}.
Among these approaches, probability theory is the most commonly used method in visual analytics.
For example, Correa~\cite{Correa2009_VAST} and Wu et al.~\cite{Wu2012_InfoVis} regarded uncertainty as a parameter that describes the dispersion of measured values.
Specifically, they represented uncertainty as an estimated standard deviation, in which the measured value is defined on the set of both positive and negative real numbers.
Since the measured value (the ranking score) in our approach is defined on the set of positive real numbers, the above modeling method cannot be directly applied to our work.
Therefore, we employ a Poisson mixture to model uncertainty.

\section{MutualRanker Overview}

\subsection{Requirement Analysis}
\label{sec:requirments}
The research problems were gradually identified in our own research projects related to Twitter data analysis.
In these projects, we often \doc{needed} to discover and retrieve relevant tweets, users, and hashtags by keyword search.
\shimei{Frequently,} we \shimei{also} \doc{needed} to manually check the data and improve the quality using heuristics.
\shimei{This} process \shimei{can be very} time-consuming and requires domain expertise.

\revision{To address this issue, we collaborated with two domain experts to develop MutualRanker, including one researcher in sociology (S) and one researcher in media and \doc{communications} (C).}
The experts are experienced in retrieving data from microblogs.
They also \shimei{had experience using} a method similar to the one described \shimei{above.}
We conducted several \shimei{interviews with them}, mainly focusing on probing their needs and microblog retrieval process.
We then identified the following high-level requirements based on \shimei{their} feedback.

\noindent\textbf{\normalsize R1 - Examining an initial set of salient microblog data.}
Both experts expressed the need for a ranking list of keyword search results. Keyword-based microblog retrieval results often include millions of posts and tens of thousands of users and hashtags.
Thus, these results are too massive for analysts to quickly discover relevant \shimei{data.}
The experts usually have to examine the data carefully and design a set of rules to filter out irrelevant data. As a result, they stated the need for a toolkit that can rank extracted posts, users, and hashtags to facilitate their data retrieval tasks.
This need is consistent with the findings of previous research~\cite{Duan2012_coling,Weng2010_WSDM}.

\noindent\textbf{\normalsize R2 - Revealing relationships \doc{within} microblog data.} Previous \doc{research}~\cite{Chen2013,Duan2012_coling, Liu2014_VAST} has \doc{also} indicated that the relationships \doc{within} data help users locate interesting information \doc{more} easily.
Furthermore, the relationships among the three dimensions of microblog data (posts, users, and hashtags) can assist them in extracting salient data.
For example, posts from opinion leaders are usually more important than those from average users. The domain experts desired the ability to explore different types of relationships.

\noindent\textbf{\normalsize R3 - Exploring salient microblog data from different perspectives.}
Since the three \shimei{dimensions of microblog data} usually influence each other, the experts wanted to understand \doc{this} influence so that they can link important data in one dimension to that in another dimension.
For example, expert S said that, ``Collecting relevant tweets is very important for some of our projects.
After finding one important tweet, I usually check other tweets \doc{from} the same author as well as the tweets marked by the same hashtag(s). This helps me \doc{find relevant} tweets quickly.''

\noindent\textbf{\normalsize R4 - Understanding the error produced by the ranking mechanism.} The microblog data ranking mechanism is not perfect and often introduces error\shimei{s} or uncertainty into the retrieval process.
Thus, the degree of uncertainty must be analyzed and understood to facilitate informed decision-making~\cite{Correa2009_VAST,Skeels2010revealing,Wu2012_InfoVis}.
The experts requested to know which ranking scores are more error-prone.

\noindent\textbf{\normalsize R5 - Analyzing the influence of \shimei{the} error\shimei{s} of one item on other items.} The experts also expressed the need to understand error propagation among data items.
They claimed that this information can \shimei{help them} considerably in filtering out irrelevant data.
For example, expert C commented, ``When I find an item with \shimei{an incorrect} ranking score, I also want to know which items are influenced by this so that I can adjust the ranking score quickly.''\looseness=-1

\subsection{System Overview}
\begin{figure}[t]
\centering
\includegraphics[width=0.45\textwidth]{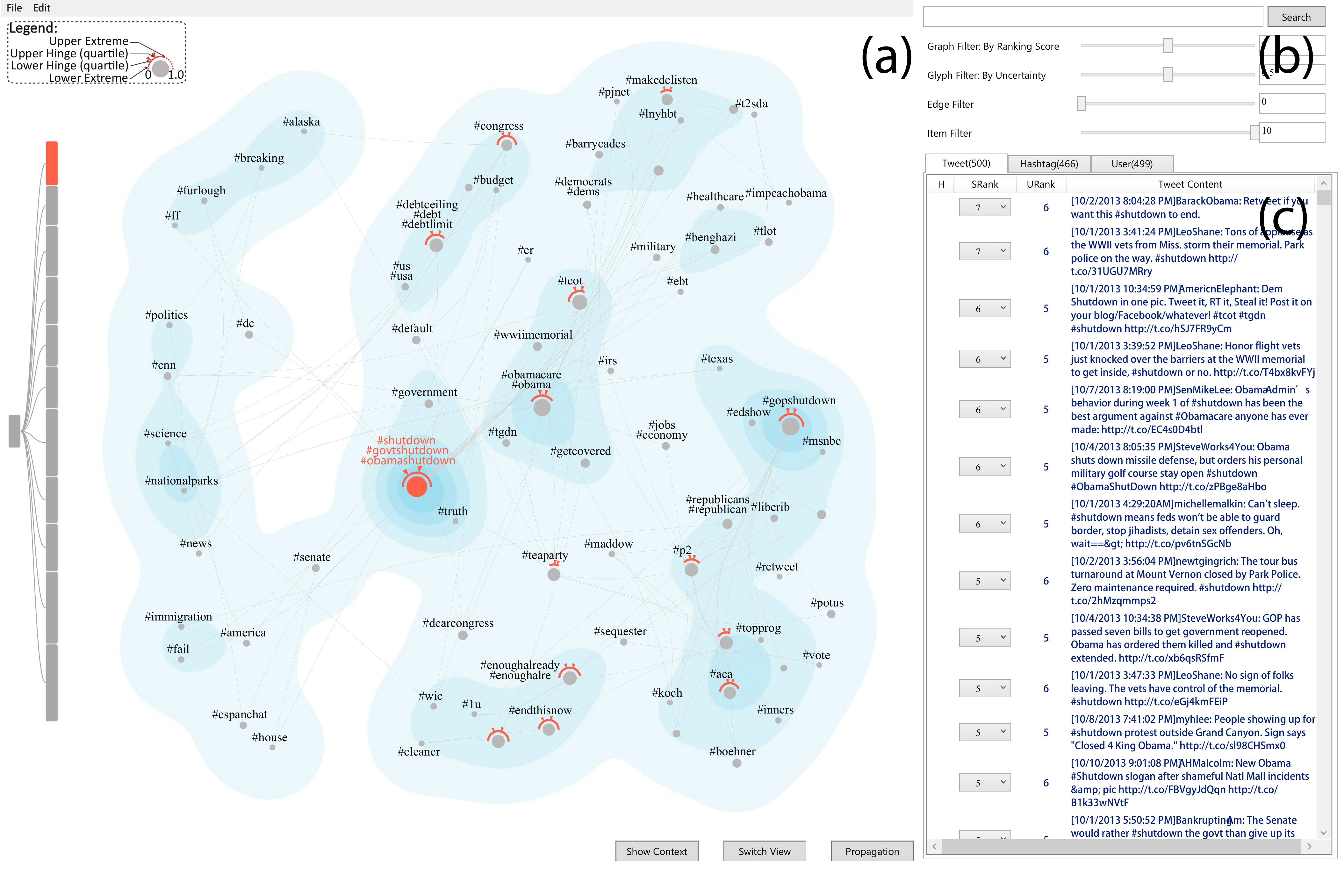}
\vspace{-2mm}
\caption{User interface: (a) MutualRanker visualization; (b) control panel; (c) information panel.}
\label{fig:userinterface}
\vspace{-5mm}
\end{figure}

The collected requirements \shimei{have motivated} us to develop a visual analytics toolkit, MutualRanker. It consists of the following components\doc{:}
\begin{compactitem}
\item \doc{An} MRG model to generate the initial ranking lists of posts, users, and hashtags (\textbf{\normalsize R1});
\item \doc{An} uncertainty model to estimate uncertainty and its topological propagation on a graph (\textbf{\normalsize R4}, \textbf{\normalsize R5});
\item A composite visualization to \shimei{present} the graph-based ranking results, uncertainty, and its propagation (\textbf{\normalsize R2}, \textbf{\normalsize R3}).
\end{compactitem}

The primary goal of MutualRanker is to extract a list of \emph{k} microblog posts/users/hashtags that are relevant \doc{to query} \emph{q}.
Fig.~\ref{fig:overview} illustrates the main components needed to achieve this goal.
Given a microblog \shimei{dataset extracted} by a query, the preprocessing module first extracts the post graph, the user graph, and the hashtag graph.
The three graphs are then fed to the MRG model, which produces three ranking lists of posts, users, and hashtags.
The uncertainty module estimates the uncertainty in the retrieval model and its topological propagation.
The visualization module takes the ranking results and the uncertainty estimation as input and illustrates them in a composite visualization that \shimei{includes} a graph visualization, an uncertainty glyph, and a flow map.
Users can interact with the generated visualization for further analysis.
For example, \shimei{a} user can modify \shimei{a} ranking result\shimei{. With this input,} MutualRanker will incrementally update the ranking results.

Fig.~\ref{fig:userinterface} depicts the user interface of MutualRanker.
It contains three different interaction areas: MutualRanker visualization (Fig.~\ref{fig:userinterface}(a)), control panel (Fig.~\ref{fig:userinterface}(b)), and information panel (Fig.~\ref{fig:userinterface}(c)).
The visualization view consists of two parts: 1) the stacked tree visualization that shows the hierarchical structure of microblog data; 2) the composite visualization that \revision{simultaneously reveal \doc{the} retrieved microblog data}, \doc{the} uncertainty of the ranking results\doc{,} and its topological propagation.
The control panel consists of a set of controls that enable users to interactively update the ranking.
 The information panel displays the corresponding microblog data such as posts, users, and hashtags for a selected aggregate item.
%
%

\section{Mutual Reinforcement Graph}
The \kgsedit{main} feature of \kgsedit{MRG}~\cite{Duan2012_coling,Wei2008_SIGIR} is that it employs both the relationships within posts, users, or hashtags\doc{,} and the relationships \kgsedit{between them to improve rankings}.
This feature significantly reduces the workload of analysts \shimei{when interacting with} our visual analytics system.
For example, if \kgsedit{an} analyst modifies the ranking score of a hashtag, MRG not only incrementally updates the ranking scores of the neighboring \kgsedit{hashtags}, but also those \kgsedit{of} relevant users and \shimei{posts.}
This process \shimei{allows our system to integrate} \doc{user} knowledge into the visual analytics process with acceptable \doc{user} effort.
\shimei{This is also the main reason why} we adopt MRG in MutualRanker.

\begin{figure}[h]
  \centering
  \includegraphics[width=2in]{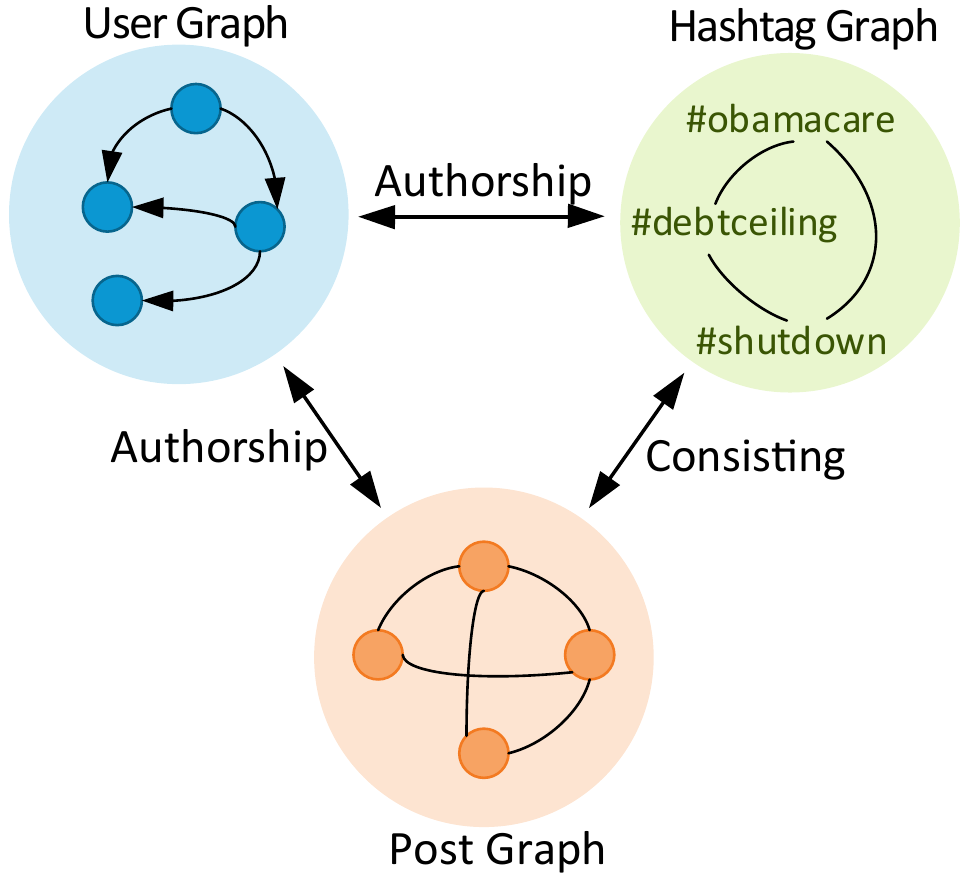}
  \vspace{-2mm}
  \caption{
  Mutual reinforcement model.}

  \label{fig:mrf_model}
  \vspace{-3mm}
\end{figure}




The input of MRG includes three graphs, the post graph, the user graph\doc{,} and the hashtag graph, \shimei{as well as} the relationships among \shimei{them}.
The three graphs and their relationships are shown in \kgsedit{Fig.}~\ref{fig:mrf_model}.
As in~\cite{Duan2012_coling}, the post graph is built based \kgsedit{on} cosine similarity.
\revision{
A recent study has shown that cosine similarity with \doc{a} TFIDF weighting scheme is the most appropriate measure to compute the similarity between microblog posts~\cite{sedhai2014hashtag, Zangerle2013impact}.
As a result, we employ cosine similarity in our system.
}
The user graph is constructed based \kgsedit{on} follower-followee relationships.
The hashtag graph is \kgsedit{generated} according to the co-occurrence of two hashtags.
\kgsedit{The} three graphs are \kgsedit{also} connected by two relationships: authorship and co-occurrence.
If a user publishes a post, \kgsedit{then} we connect this user with \kgsedit{his/her} post.
We also link this user with all \kgsedit{of} the hashtags in this post.
\shimei{Each} post is \shimei{also} linked to all of the hashtags associated with it.\looseness=-1

For simplicity, we uniformly denote posts, users, and hashtags as items in the following discussion.

The MRG employs a method similar to PageRank~\cite{Brin1998anatomy} to model the mutual influence among different items in heterogeneous graphs:

\begin{equation}
\small
\label{eq:4.1}
\left[ \begin{matrix}
   {{R}_{p}}  \\
   {{R}_{u}}  \\
   {{R}_{h}}  \\
\end{matrix} \right]=d\left[ \begin{matrix}
   {{\alpha }_{pp}}{{M}_{pp}} & {{\alpha }_{up}}{{M}_{up}} & {{\alpha }_{hp}}{{M}_{hp}}  \\
   {{\alpha }_{pu}}{{M}_{pu}} & {{\alpha }_{uu}}{{M}_{uu}} & {{\alpha }_{hu}}{{M}_{hu}}  \\
   {{\alpha }_{ph}}{{M}_{ph}} & {{\alpha }_{uh}}{{M}_{uh}} & {{\alpha }_{hh}}{{M}_{hh}}  \\
\end{matrix} \right]\left[ \begin{matrix}
   {{R}_{p}}  \\
   {{R}_{u}}  \\
   {{R}_{h}}  \\
\end{matrix} \right]+(1-d)\left[ \begin{matrix}
   {{W}_{p}}  \\
   {{W}_{u}}  \\
   {{W}_{h}}  \\
\end{matrix} \right].
\end{equation}

$R_p$, $R_u$, and $R_h$ are the ranking score vectors of posts (p), users (u), and hashtags (h). $M_{xy}$ denotes the affinity matrix from $x$ to \kgsedit{$y$,} where $x,y$ can be posts, users, or hashtags.
${\alpha}_{xy}$ is a weight used to balance the mutual reinforcement strength \kgsedit{among} posts, users, and hashtags.
$d$ is the damping factor in \kgsedit{PageRank,} and \kgsedit{we} set it to 0.85\doc{,} as in~\cite{Brin1998anatomy}. $W_p$, $W_u$, and $W_h$ are \kgsedit{vectors for the} prior saliency of the items (e.g., the content quality of posts, the social influence of users, or the popularity of \doc{hashtags).}

Let
$\begin{small} R=\left[ \begin{matrix}
{{R}_{p}}  \\
{{R}_{u}}  \\
{{R}_{h}}  \\
\end{matrix} \right] \end{small}$
,
$\begin{small} W=\left[ \begin{matrix}
   {{W}_{p}}  \\
   {{W}_{u}}  \\
   {{W}_{h}}  \\
\end{matrix} \right] \end{small}$
, and
$\begin{small} M=\left[ \begin{matrix}
   {{\alpha }_{pp}}{{M}_{pp}} & {{\alpha }_{up}}{{M}_{up}} & {{\alpha }_{hp}}{{M}_{hp}}  \\
   {{\alpha }_{pu}}{{M}_{pu}} & {{\alpha }_{uu}}{{M}_{uu}} & {{\alpha }_{hu}}{{M}_{hu}}  \\
   {{\alpha }_{ph}}{{M}_{ph}} & {{\alpha }_{uh}}{{M}_{uh}} & {{\alpha }_{hh}}{{M}_{hh}}  \\
\end{matrix} \right] \end{small}$
\kgsedit{. Then,} Eq.~(\ref{eq:4.1}) can be simplified as:

\begin{equation}
\small
\label{eq:4.2}
R=dMR+(1-d)W.
\end{equation}


%
%
%
%
%
%
%
%
\section{MRG-Based Uncertainty Analysis}
Since exact inference of MRG is very time-consuming on a large graph, we approximate \shimei{it} using \shimei{a more efficient} Monte Carlo sampling \shimei{method.}
\shimei{We also} explicitly model the uncertainty associated with each item (e.g., a post, a user\doc{,} or a hashtag)\kgsedit{, as} well as its propagation on the graph.


\subsection{MRG Computation with Monte Carlo Sampling Method}

Duan et al.~\cite{Duan2012_coling} proposed a matrix-based method to solve MRG, which iteratively updates the ranking scores using Eq.~(\ref{eq:4.1}).
The matrix-based method is a global \doc{one}.
An update to any item is achieved by running the method on the \shimei{entire} item set, which is very \shimei{time-consuming.}
To \kgsedit{address} this problem, we use the Monte Carlo sampling \kgsedit{method}.
\kgsedit{The} advantages of \kgsedit{this} sampling method \kgsedit{over the matrix-based method are as follows}~\cite{Avrachenkov2007}:
\begin{compactitem}
\item \kgsedit{Ranking} scores \kgsedit{are locally updated} when the input changes locally;
\item \kgsedit{The} ranking scores of important items \kgsedit{are accurately estimated} after a few iterations;
\item \kgsedit{The} uncertainty of the ranking scores \kgsedit{are modeled} accurately because \kgsedit{the Monte Carlo} method calculates variance statistically.
\end{compactitem}

To employ \kgsedit{this} \kgsedit{method}, we first solve $R$ \kgsedit{as formulated} by Eq.~(\ref{eq:4.2}):



\begin{equation}
\small
\label{eq:5.1.1}
R  = (1 - d)(I - dM)^{ - 1}W  = (1 - d)(\sum\limits_{k = 0}^\infty  {d^k M^k })W.
\end{equation}

\revision{We then perform a series of random walks for each item.}
A random walk may stop at each step with a probability of $1 - d$.
If \kgsedit{the walk} continues, \kgsedit{then it proceeds} to the next step according to the matrix $M$.
Each element $m_{ij} $ defines the transition probability from $i$ to $j$.

Let \begin{small}$Z = \sum\limits_{k = 0}^\infty  {d^k M^k }$\end{small}\kgsedit{. The} ranking score of each item is:
\begin{equation}
\small
\label{eq:5.1.2}
r_j  = (1 - d)\sum\limits_{i = 1}^N {w_i z_{ij} },
\end{equation}
where the element $z_{ij} $ in $Z$ is the average number of times that \shimei{a} random walk starting from item $i$ visits item $j$.
We estimate $z_{ij}$ by computing the empirical mean of a number of random walks.

Duan et al.~\cite{Duan2012_coling} \kgsedit{only consider} the similarity of items \kgsedit{in computing} $m_{ij}$.
\kgsedit{Thus, high-ranking} scores \kgsedit{may be incorrectly assigned to} users who \doc{publish} \kgsedit{many} posts \kgsedit{that do not receive} attention.
To \kgsedit{address} this, we \kgsedit{consider the} prior saliency of \kgsedit{items} in the sampling \kgsedit{process}.
Specifically, the transition probability from $m_{ij}$ is defined as $similarity(i,j) \cdot w_j$.

\subsection{Uncertainty Modeling}

\revision{In MutualRanker, we use an approximation \shimei{method} to solve MRG, which may introduce uncertainty into the retrieval result\doc{s}.
It is therefore important to model uncertainty.
Since we employ the Monte Carlo sampling method, the distribution of each ranking score is \shimei{known.}
\shimei{Hence,} we \shimei{can} employ \shimei{the} probability theory to model uncertainty.}\looseness=-1

\doc{Uncertainty is defined as a parameter for depicting} the dispersion of \kgsedit{values} that can \kgsedit{be reasonably} attributed to the measured value~\cite{BIPM1995}.
Traditional methods model the measured value as a normally distributed random variable~\cite{Correa2009_VAST,Wu2012_InfoVis}.
Variance~\cite{Correa2009_VAST} and standard deviation~\cite{Wu2012_InfoVis} are \kgsedit{among the most} commonly used measures to represent \kgsedit{uncertainty wherein} the measured value is defined on the set of both positive and negative real numbers.

The measured value (\kgsedit{ranking} score) in our approach is defined on the set of positive real numbers\doc{. Thus, }the above modeling method cannot be \kgsedit{applied directly} to our work.

\revision{According to~\cite{Avrachenkov2007}, $z_{ij}$ in Eq.~(\ref{eq:5.1.2}) has a Poisson distribution.}
The ranking score is the weighted sum of a series of $z_{ij} $.
\kgsedit{Hence}, the ranking score is modeled as a Poisson mixture.
For \kgsedit{a} Poisson \kgsedit{mixture,} the variance is approximately proportional to the \kgsedit{mean. Hence, if} we use variance to model uncertainty, the larger \kgsedit{the} ranking \kgsedit{score}, the more uncertain it is, \kgsedit{but this is} not always true.

Standard deviation is the square root of variance and has \kgsedit{a} similar problem.
Consequently, variance and standard deviation are not good measures \doc{for depicting} uncertainty in our \shimei{model}.

For such \shimei{a} distribution, a commonly used measure \shimei{of} dispersion is the variance-mean-ratio (VMR)~\cite{Cox1966}. The higher the \kgsedit{VMR}, the more dispersed \kgsedit{the distribution}. For item $j$, its VMR ($u_j$) can be defined as:\looseness=-1

\begin{equation}
\small
\label{eq:5.2.1}
u_j  = {{v_j }}/{{r_j }} ,
\end{equation}
where $v_j $ is the distribution variance of the ranking score of item $j$.
According to~\cite{Avrachenkov2007}, $v_j$ can be calculated as follows:

\begin{equation}
\small
v_j  = (1 - d)^2 \sum\limits_{i = 1}^N {w_i^2 v_{z_{ij}}}.
\end{equation}
where $v_{z_{ij}}$ is the variance of ${z_{ij}}$. \kgsedit{Each} $z_{ij}$ obeys a Poisson distribution \kgsedit{and its} variance can be calculated from its expectation.

The \shimei{massive number} of items in the microblog data \kgsedit{means} we cannot place all of them on the screen.
\kgsedit{Hence, we} aggregate similar items to form a cluster.
The overall ranking score of a \doc{cluster $r_c$} is defined as the sum of \shimei{the} ranking scores of its items~\cite{Bianchini2005}.
The ranking scores are independent \kgsedit{of} each \doc{other and} the overall variance of the cluster, $v_c$, is the sum of the variance of the ranking scores. \kgsedit{Thus,} the uncertainty of a cluster, $u_c$, can be \kgsedit{calculated naturally} by dividing $v_c$ and $r_c$.

\begin{equation}
\small
\label{eq:5.2.2}
{{u}_{c}}={{{v}_{c}}}/{{{r}_{c}}}=\sum\limits_{j\in c}{(r_j/r_c)(v_j/r_j)}=\sum\limits_{j\in c}{{{k}_{j}}{{u}_{j}}}.
\end{equation}
\kgsedit{Eq.~(\ref{eq:5.2.2}) shows that} $u_c$ can be expressed by \doc{a} weighted sum of the uncertainty of its items where each weight $k_j$ is the ratio of the ranking scores of item $j$ and \kgsedit{cluster} $c$.
\kgsedit{Thus,} the uncertainty of a cluster is \kgsedit{mainly determined} by its important items.

\subsection{Topological Uncertainty Propagation}
\label{sec:propagation}

\revision{If an analyst finds an \shimei{incorrectly ranked item}, he can modify it based on his knowledge.
He can further track how the uncertainty propagates from one cluster to another \shimei{to identify other affected} items.}
To help \shimei{an} analyst track uncertainty, we explicitly model its topological propagation on the graph.

In MRG, the ranking score of an item can be expressed as a linear combination \kgsedit{of} ranking scores of related items.
\kgsedit{Hence}, the variances of a ranking score can also be expressed as a linear combination of the variances \kgsedit{of} related ranking scores.
\kgsedit{The} uncertainty of each item can be calculated from its ranking score and its variance, \kgsedit{and hence,} the uncertainty of an item can also be \kgsedit{expressed linearly by the} uncertainty of other items. \kgsedit{Specifically,}

\begin{equation}
\small
\label{eq:5.3.1}
u_j  = \sum\limits_{i = 1,i \ne j}^N {m_{ij}^* u_i },
\end{equation}
where each $m_{ij}^* = ({d^2 m_{ij}^2 r_i })/({(1 - d^2 m_{jj} )r_j })$. \kgsedit{Eq.~(\ref{eq:5.3.1}) shows that the} uncertainty of each item is not independent and \shimei{it} \kgsedit{propagates} on the graph in a linear form.
\kgsedit{Thus,} for each pair of items $i$ and $j$, $m_{ij}^* u_i $ can be viewed as the propagated uncertainty from item $i$ \kgsedit{to} $j$. We denote it by $u_{i \to j}$.

Rewriting Eq.~(\ref{eq:5.3.1}) in a matrix form, we can formulate the uncertainty propagation as a Markov chain:

\begin{equation}
\small
UM^*  = U,
\end{equation}
where $U = [u_j]_{1 \times N}$ and $M^* = [m_{ij}^*]_{N \times N}$.

Similar to the uncertainty propagation from item to item, we can model the uncertainty propagation from cluster to cluster using the following procedure.
First, based on Eq.~(\ref{eq:5.3.1}), we calculate the propagated uncertainty from each item $i$ in the source cluster $c_s$ to each item $j$ in the target cluster $c_t$ (Fig.~\ref{fig:uncertainty_2}(a)).
Second, for each item $j$ in $c_t$, we compute the propagated uncertainty $u_{c_s \to j}$ from $c_s$ to item $j$ by aggregating the uncertainty propagated from each item in the source cluster (Fig.~\ref{fig:uncertainty_2}(b)).

\begin{equation}
\small
u_{c_s \to j}  = \sum\limits_{i \in c_s } {u_{i \to j} }.
\end{equation}
Finally, the uncertainty of a cluster is a weighted sum of the uncertainty of the items in it (Eq.~(\ref{eq:5.2.2})).
\kgsedit{Thus,} the overall propagated uncertainty $u_{c_s \to c_t}$ from $c_s$ to $c_t$ can be calculated as \kgsedit{the} weighted sum of the propagated uncertainty from $c_s$ to each $j$ in $c_t$ (Fig.~\ref{fig:uncertainty_2}(c)).



\begin{equation}
\small
u_{c_s \to c_t}  = \sum\limits_{j \in c_t } {k_j u_{c_s \to j} }.
\end{equation}

\begin{figure}[t]
  \centering
  \includegraphics[width=0.45\textwidth]{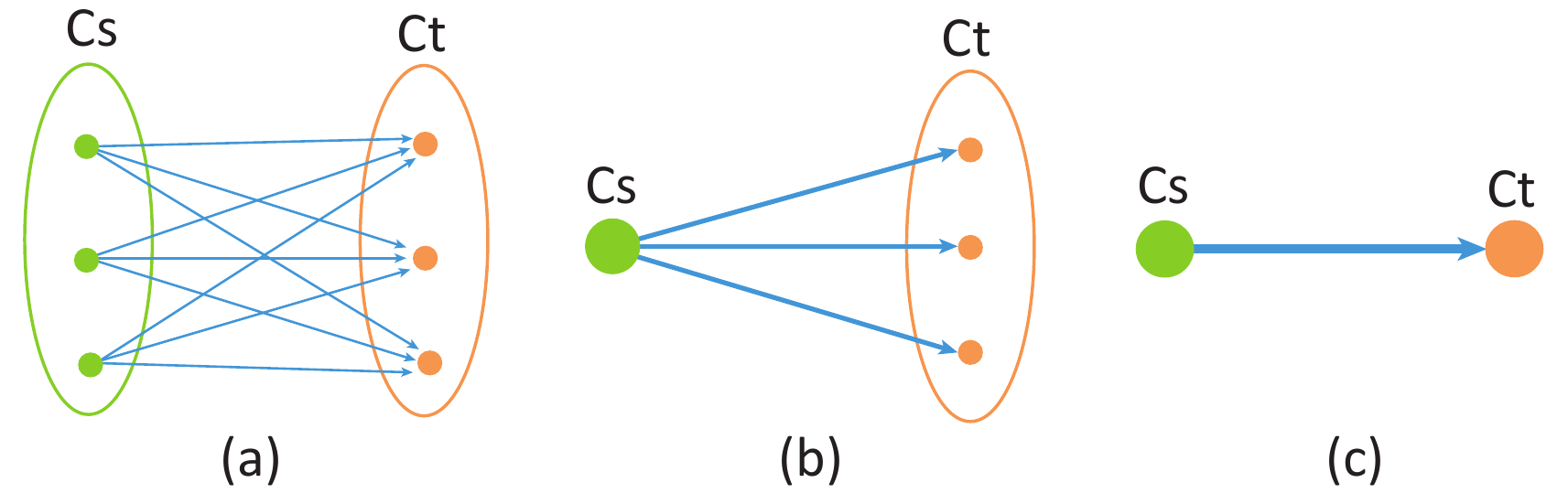}
  \vspace{-3mm}
  \caption{
  Topological uncertainty propagation calculation}
  \label{fig:uncertainty_2}
  \vspace{-5mm}
\end{figure}


\subsection{Incremental Ranking Update}

We also allow analysts to interactively modify the item ranking result based on their knowledge.
\kgsedit{We can update the model locally because we} use the Monte Carlo sampling method.
\kgsedit{After} the analyst changes the ranking score(s), our approach iteratively updates the prior salience \shimei{score(s)} of the item(s).
Accordingly, the affinity matrix is changed from $M$ to $M'$.
This change \shimei{only} affects a small part of \kgsedit{the} random walks \kgsedit{used in the} Monte Carlo sampling method.

For \kgsedit{the} affected random walks, existing incremental graph ranking algorithms ~\cite{Bahmani2010} perform re-sampling and update the ranking scores by aggregating the statistics of these new random walks into the original \kgsedit{results}.
\shimei{One} main problem \doc{with} these algorithms is that re-sampling \shimei{requires} \doc{a} considerable \shimei{amount of} time\doc{,} \shimei{which may make} real-time interaction \shimei{impossible}.
Suppose $n$ is the average number of neighbors that an item \shimei{has} and $l$ is the average length of a sampled random walk.
At each step in a random walk, we have to sample from a multinomial distribution with $n$ possible outcomes and the time cost is $O(n)$.
Thus\doc{,} sampling a new random walk will take $O(nl)$ time.
\kgsedit{The} time needed \kgsedit{to compute and aggregate the} statistics of \kgsedit{these} samples is $O(l)$.
\doc{The} total time \shimei{required} for a new sample is $O(nl) + O(l)=O(nl)$.

However, in our scenario, we do not \kgsedit{delete} or add edges on the graphs.
As a result, we do not need to perform re-sampling.
\shimei{We} only \shimei{need to} modify the statistics of a random walk based on the modified transition probability $m_{ij}$, thereby avoiding the high cost \shimei{associated with resampling.}
The time cost of updating an influenced random walk is reduced to $O(l)$.



Given a random walk: $path = \{ i \to n_1  \to ... \to n_{k - 1}  \to j\}$, we define a new random variable $x_{ij}^{k} $.
In particular, $x_{ij}^{k}  = 1$ indicates \kgsedit{that} the random walk \kgsedit{starts} from $i$ and \kgsedit{reaches} $j$ by moving $k$ steps.
The original weight of each step in the random walk is $1$.
During \kgsedit{an} update, we re-calculate the weight of this step \shimei{using} ${{P'(x_{ij}^{k}  = 1)}}/{{P(x_{ij}^{k}  = 1)}}$.
$P(x_{ij}^{k}  = 1)$ is the probability of $x_{ij}^{k}  = 1$ according to $M$ and $P'(x_{ij}^{k}  = 1)$ is the probability of $x_{ij}^{k}  = 1$ according to $M'$.
\kgsedit{Hence,} $P(x_{ij}^{k}  = 1)$ is calculated by:

\begin{equation}
\small
\label{eq:5.4.1}
P(x_{ij}^{k}  = 1) = m_{in_1 } m_{n_1 n_2 } ...m_{n_{k - 1} j}.
\end{equation}
Similarly, $P'(x_{ij}^{k}  = 1)$ can also be calculated. 

\section{Visualization}\label{sec:vis}
To help analysts extract microblog data of interest interactively, we \doc{have designed} a composite visualization \doc{that includes} a graph visualization, an uncertainty glyph, and a flow map (Fig~\ref{fig:shutdownoverview}(a)).

\subsection{Ranking Results as Graph Visualization}

\begin{figure*}[t]
\centering
\includegraphics[width=0.95\textwidth]{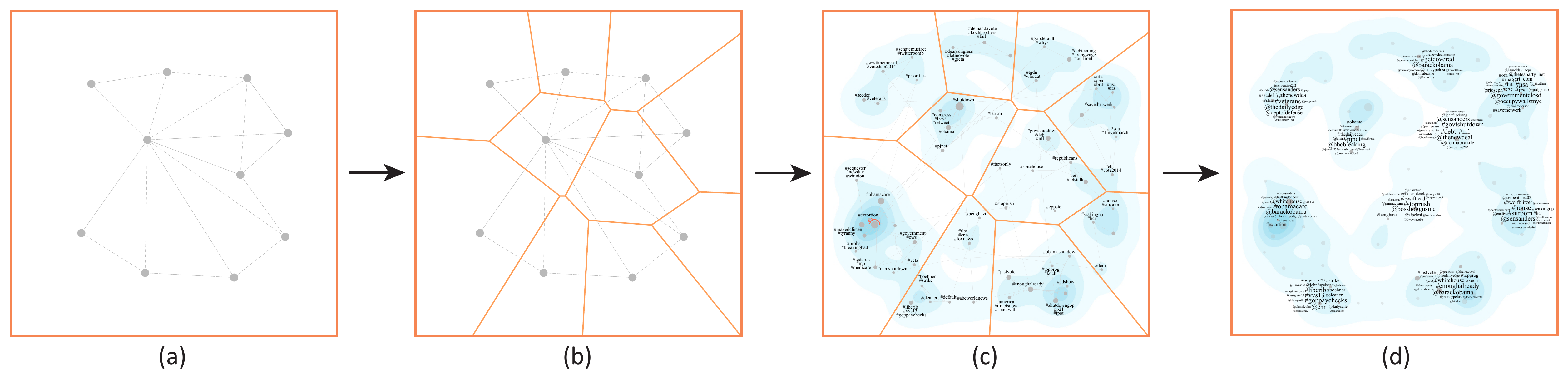}
\vspace{-3mm}
\caption{\kgsedit{Basic} idea of the layout algorithm: (a) place the cluster nodes and derive the layout center of each cluster; (b) compute the Voronoi tessellation and treat the cell as the layout area of each cluster; (c) \kgsedit{layout} of \kgsedit{representative} and non-representative nodes; \kgsedit{and }(d) final layout result with context.}
\vspace{-4mm}
\label{fig:layout}
\end{figure*}
\revision{\shimei{Since one} post corresponds to only one user and a few hashtags, the scope of influence of a post is smaller than that of a user or a hashtag.
Updating the ranking score of a post will only directly affect the ranking scores of its author, a few related hashtags, and a number of posts.
In contrast, updating the ranking score of a user or hashtag will directly impact the ranking scores of hundreds or even thousands of posts as well as a number of users and hashtags.
On the other hand, the number of posts is usually huge\doc{,} around 10-100 times that of users or hashtags.}
\kgsedit{Analysts would require more time} to provide their feedback on a post graph.
As a result, we regard the user and the hashtag as the primary visualization elements and the post as \doc{a} secondary \doc{element} mainly used to illustrate the content of the primary elements.
Accordingly, users and hashtags are \kgsedit{visually represented} by a node-link graph \kgsedit{whereas posts are represented as} a list.
For simplicity, we take a hashtag graph as an example to illustrate the basic idea \kgsedit{of} graph visualization.\looseness=-1

To allow analysts to navigate large graphs efficiently, a hierarchy is built based on \doc{a} Bayesian Rose Tree~\cite{Liu2012} with each non-leaf node representing a hashtag cluster.
As shown in Fig.~\ref{fig:userinterface}(a), a stacked tree is adopted to represent the hashtag hierarchy and a density-based graph visualization is employed to illustrate the relationships within the user/hashtag graphs and between them (\textbf{\normalsize R2}, \textbf{\normalsize R3}).

\revision{The density-based graph visualization combines a node-link diagram with a density map} to display the nodes at the selected level of the hashtag tree.
As in~\cite{Liu2014_VAST}, we extract representative nodes for each of the cluster nodes at the selected tree level and assign other non-representative nodes to their closest representative nodes.
As shown in Fig.~\ref{fig:shutdownoverview}(a), the representative nodes are displayed as a node-link diagram and the other nodes as a density map.
In this visualization, the representative nodes of one cluster are placed near each other to reflect their closeness.
\revision{The size of the node encodes the sum of the ranking score of each item.}
The corresponding users \kgsedit{are} overlaid around the selected hashtag node to provide more analysis context (Fig.~\ref{fig:layout}(d)).

\noindent\textbf{\normalsize Layout.} The layout of the stacked tree is quite straightforward.
\kgsedit{Thus, we} introduce the layout of the density-based graph, which contains the following steps.

\noindent\emph{\normalsize Step 1: \kgsedit{Derive the} layout center of each cluster at the selected tree level}.
\kgsedit{We} build a cluster graph by checking the edge connections between the two cluster nodes.
An edge is added if \doc{a} sufficient \doc{number of} connections between the two cluster nodes \kgsedit{can be found}. \kgsedit{The} cluster graph is \kgsedit{then} placed by a force-directed layout~\cite{Kamada1989}.
As shown in Fig.~\ref{fig:layout} (a), the position of each cluster node is treated as the center of each hashtag cluster.

\noindent\emph{\normalsize Step 2: \kgsedit{Compute} the layout area of each cluster.}
In this step, we compute the corresponding Voronoi tessellation based on the cluster center.
The corresponding tessellation cells are treated \kgsedit{as} layout areas of \kgsedit{the} hashtag clusters (Fig.~\ref{fig:layout}(b)).

\noindent\emph{\normalsize Step 3: Layout of \kgsedit{representative} and non-representative nodes.}
In this step, the force-directed layout is adopted to place the representative nodes.
To ensure the representative nodes within one cluster are placed in the corresponding cluster layout area, a repulsion force is added from the area boundary to each node within this area.
The kernel density estimation~\cite{Lampe2011} is utilized to represent the distribution of non-representative nodes (Fig.~\ref{fig:layout}(c)).

\noindent\emph{\normalsize Step 4: \kgsedit{Layout} of the context word cloud.}
\revision{\doc{Showing} the hashtag graph and user graph simultaneously \doc{would introduce visual clutter}.
To solve this issue, we treat the hashtag graph as \doc{a} primary \doc{element} and the user information as context.}
In particular,
when a hashtag node is selected, a word cloud that includes the users who use this hashtag is laid out to \kgsedit{pro}vide user context.
In this word cloud, the selected hashtag is placed in the middle.
\kgsedit{A} sweep-line-based word cloud layout algorithm~\cite{Shi2010_VAST} is employed \kgsedit{to produce such a word cloud}.
Fig.~\ref{fig:layout} (d) shows a layout result with a word cloud context.

\noindent\textbf{\normalsize Interaction.} The following interactions are provided to \kgsedit{assist} analysts \kgsedit{in investigating} the ranking results from multiple perspectives.

\noindent\emph{\normalsize Examining the ranked microblog data and their relationships} (\textbf{\normalsize R2}).
The density-based graph visualization \doc{provides} an easy way to explore the ranking results \kgsedit{from} the hashtag or user perspective.
\kgsedit{Utilizing} the hashtag hierarchy \kgsedit{allows} the analyst \kgsedit{to} explore the ranking \kgsedit{results} from a global overview to local details.
Several \kgsedit{filters,} such as the edge \kgsedit{or the} glyph \kgsedit{filter,} enable analysts \kgsedit{to} customize this view \kgsedit{easily}.
\kgsedit{Relevant} posts, hashtags, and users are also provided to help analysts better understand \doc{the} content of the \doc{selected} cluster node.

\noindent\emph{\normalsize Smoothly switching between different data dimensions} (\textbf{\normalsize R3}).
Inspired by the context popup interaction in~\cite{Ghani2013_TVCG}, we also overlay context of a selected item to \kgsedit{provide} further navigation \kgsedit{cues}.
For example, if the analyst selects a hashtag, the labels of users who use that hashtag can be overlaid around the selected hashtag via a word cloud (Fig.~\ref{fig:usercontext}(a)).
If the analyst finds something \kgsedit{of} interest, the \revision{hashtag graph} will be smoothly \doc{transitioned} to the \revision{user graph} (Fig.~\ref{fig:usercontext}(b)).

\subsection{Uncertainty as Glyph}


After testing with the first prototype, the experts identified several incorrect ranking results.
\doc{They} expressed the need to be informed \kgsedit{of such} results.
This requirement is related \kgsedit{intimately with the} conclusion of previous work, which stated that effectively conveying uncertainty is very important to the visual analytics process~\cite{Correa2009_VAST,Wu2012_InfoVis}.
Since the ranking results are aggregated into clusters in the overview, the experts wanted to examine the uncertainty distribution of the aggregate node, including \revision {the minimum value (0), maximum value (1.0)}, lower extreme, upper extreme, lower hinge (25\%), and upper hinge (75\%).

Inspired by the box plot design (Fig.~\ref{fig:glyph}(a)), we \doc{have designed} a glyph to meet the above \doc{requirements} (Fig.~\ref{fig:glyph}(b)).
As shown in Fig.~\ref{fig:glyph}(a), \revision{six values from a \doc{set of data} are conventionally used in a box plot, including the minimum and maximum values, the extremes\doc{, and} the upper and lower hinges (quartiles).} A total of 50\% percent of items fall in between the upper and lower hinges.
To combine a box plot with a graph node, we first transform the box plot to a line-based one, and then bend it around the upper boundary of the node (Fig.~\ref{fig:glyph}(b)).
We also attempted several alternatives in the participatory design process with \kgsedit{experts}.
Fig.~\ref{fig:glyph}(c) is one of them.
After interacting with this alternative, the experts \kgsedit{stated} that it was confusing.
They thought that the item with \doc{more of} \kgsedit{a} filled area inside should \kgsedit{be} the one on which they should \kgsedit{focus}.
\kgsedit{However, in reality, these nodes were only} nodes with \kgsedit{a} larger area between the upper \kgsedit{and lower hinges}.
A PhD student from \doc{an} art school later confirmed \kgsedit{that a larger amount of} digital ink will attract more attention from users.
After several interactions with the experts and the art student, we choose Fig.~\ref{fig:glyph}(b) as our final design.

Analysts can obtain an overview of the uncertainty distribution in a cluster by examining its uncertainty glyph.
Fig.~\ref{fig:glyphtype} illustrates several example patterns.
For example, in Fig.~\ref{fig:glyphtype}(a), the majority of items in this cluster are characterized by low uncertainty. However, the cluster also contains some items with higher uncertainty.
As a result, exploring the items with high uncertainty is a worthwhile endeavor.

\begin{figure}[t]
\centering
\includegraphics[width=0.45\textwidth]{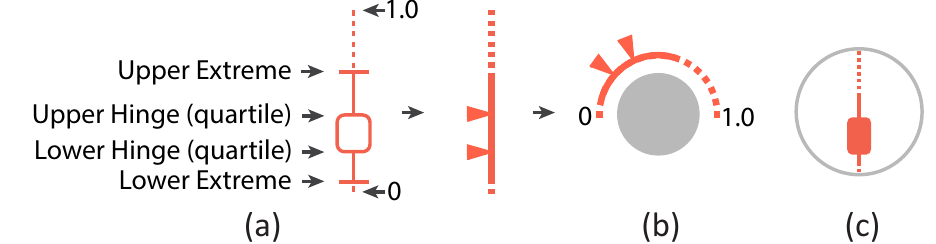}
\vspace{-1mm}
\caption{\kgsedit{Design} of the uncertainty glyph: (a) box plot; (b) transforming the box plot to the uncertainty glyph; \revision{(c)} an alternative design.}
\vspace{-4mm}
\label{fig:glyph}
\end{figure}

\noindent\textbf{\normalsize Interaction.} In addition to allowing analysts to examine the \doc{uncertainty score} (\textbf{\normalsize R4}), we also provide \kgsedit{the} interaction \kgsedit{shown below} to integrate \shimei{an} \revision{expert's} knowledge into the retrieval process.

\begin{figure}[t]
\centering
\includegraphics[width=0.45\textwidth]{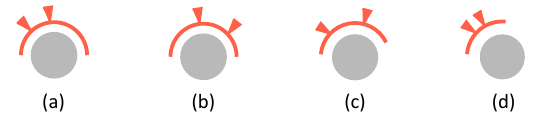}
\vspace{-2mm}
\caption{Four example patterns of the uncertainty glyph: (a) most items in this cluster have low uncertainty but some items with higher uncertainty are also included; (b) most items in this cluster have high uncertainty and some items with higher uncertainty also occur; (c) uncertainty distribution is uniform; (d) most items in this cluster have lower uncertainty.}
\vspace{-4mm}
\label{fig:glyphtype}
\end{figure}

\noindent\emph{\normalsize Interactive ranking refinement}.
After an expert finds an incorrect ranking result by examining the uncertainty glyph, \kgsedit{the expert can} \doc{modify} the ranking result.
The ranking scores of the corresponding graph nodes will also \kgsedit{be updated} accordingly.
As shown in Figs.~\ref{fig:shutdownoverview}(c)- (f), \revision{the ranking scores (e.g., node sizes) of several nodes changed.
A glyph is designed to illustrate the change, with the dotted orange circle encoding the previous ranking score and the boundary of the filled circle (gray color) representing the changed ranking score (Figs.~\ref{fig:shutdownoverview}(d)-(f)).}

\subsection{Uncertainty Propagation as Flow Map}

The flow map~\cite{Phan2005flow,Verbeek2011flow} is designed to visually analyze the movement of objects from one location to multiple locations.
Inspired by this design, we develop the uncertainty propagation path (Fig.~\ref{fig:shutdownoverview}), which is useful for quickly deriving the unknown uncertain node(s) from the known one(s) (\textbf{\normalsize R5}).

\noindent\textbf{\normalsize Layout.} The layout of multiple uncertainty propagation paths of different nodes is based on the flow map layout in~\cite{Verbeek2011flow} and the edge bundling in~\cite{Holten2009force}.
\kgsedit{The layout} contains the following steps.

\noindent\emph{\normalsize Step 1: \kgsedit{Derive} the initial uncertainty propagation path based on the flow map layout}.
We first compute the uncertainty propagation of the selected node based on the topology by using the method in Sec.~\ref{sec:propagation}.
\kgsedit{The} flow map layout via spiral trees is \kgsedit{then} unitized to generate the initial uncertainty propagation path (Fig.~\ref{fig:propagation}(a)).

\noindent\emph{\normalsize Step 2: \kgsedit{Employ} edge compatibility measures to match the corresponding propagation paths from different nodes.}
In this step, we employ the three compatibility measures described in~\cite{Holten2009force} to match the propagation paths from different nodes.

The first measure is angle compatibility, which aims to match the edges with \doc{a} smaller angle. It is defined by:
\[
\small
C_{\alpha}(e_i,e_j)=|cos(\alpha)|.
\]

The second measure is scale compatibility, which tends to match the edges with similar lengths. It is measured by:
\[
\small
C_s(e_i,e_j)=2/(l_{avg}\cdot min(|e_i|,|e_j|)+min(|e_i|,|e_j|)/l_{avg}),
\]
 where \begin{small}$l_{avg} =(|e_i|+|e_j|)/2$.\end{small}

The third measure is position compatibility, which aims to match the close edges together. It is defined by:
\[
\small
C_p(e_i,e_j)=l_{avg}/(l_{avg}+\|Q_{m1}-Q_{m2}\|),
\]
where \begin{small}$Q_{m1}$\end{small} and \begin{small}$Q_{m2}$\end{small} are the midpoints of edges $e_i$ and $e_j$.

\begin{figure}[t]
\centering
\includegraphics[width=0.45\textwidth]{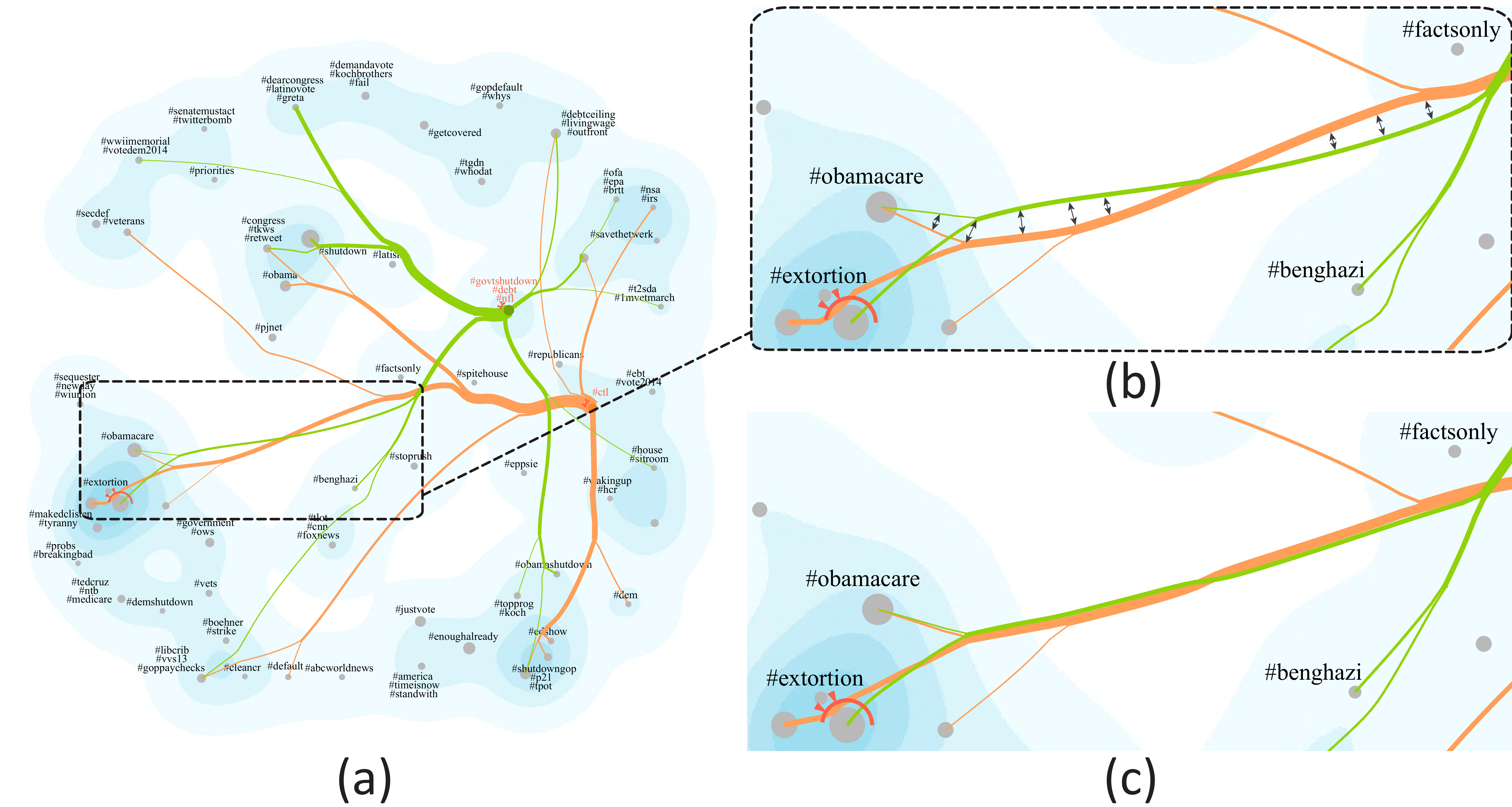}
\caption{Layout of multiple uncertainty propagation paths: (a) initial layout based on the flow map layout; (b) the matched result of the propagation paths; (c) the layout result of the propagation paths.}
\vspace{-3mm}
\label{fig:propagation}
\end{figure}

The last measure, visibility compatibility, described in~\cite{Holten2009force} is not considered in our method because there are too many line segments in the propagation path generated by the flow map layout, each of which is quite \shimei{short}.
\kgsedit{Thus, if we} consider this measure, many of \kgsedit{these line segments} will not be bundled together.

The total edge compatibility is defined by:
\[
\small
C_e(e_i,e_j)=C_{\alpha}(e_i,e_j)\cdot C_s(e_i,e_j)\cdot C_p(e_i,e_j).
\]

Fig.~\ref{fig:propagation}(b) \kgsedit{shows the matched results of the} propagation paths.

\noindent\emph{\normalsize Step 3: Compute the force to bundle the propagation path.} The combined force for a point $p_i$ on $e_i$ is defined as:
\[
\small
F_{p_i}=K_i(\|p_{i-1}-p_i\|+\|p_i-p_{i+1}\|)+\sum\limits_{e_j \in E}{\|p_i-p_j\|}\cdot C_e(e_i,e_j)，
\]
where $K_i$ is the spring constant for each segment \kgsedit{and} $E$ is the set of all the matched edges of $e_i$.
In~\cite{Holten2009force}, the last item is the electrostatic force \begin{small}$F_e=1/\|e_i-e_j\|$.\end{small}
In order to  bundle the matched paths that \doc{are located} away from each other, we replace it \doc{with} an attracting spring force.
Fig.~\ref{fig:propagation}(c) \kgsedit{shows} the layout \doc{results} of the propagation paths.

\section{Quantitative Evaluation}\label{sec:evaluation}

In this section, we quantitatively \doc{evaluate} the effectiveness of our MRG computation and incremental ranking update algorithm.

\subsection{MRG Computation}\label{sec:computation}

To evaluate the performance of our MRG computation based on the Monte Carlo sampling method, we compared it with the matrix-based method proposed in~\cite{Duan2012_coling}.
We used two Twitter datasets in the experiments: government shutdown and Ebola outbreak.
The shutdown dataset contains tweets on the 2013 US government shutdown (5,132,510 tweets from Oct. 1 to Oct. 16, 2013), which were collected by using queries such as ``shutdown.''
The Ebola dataset contains tweets on the Ebola outbreak (1,425,017 tweets from Jan. 1 to Dec. 25, 2014), which were collected by using queries such as ``ebola.''
All experiments were conducted on a PC with a 3.1GHz CPU and 16 GB RAM.\looseness=-1

\begin{table}[b]
\vspace{-3mm}
  \centering
  \centering \scalebox{0.85}{
    \begin{tabular}{|c|c|c|c|c|c|c|c|}
    \hline
    \multirow{2}[4]{*}{Dataset} & \multirow{2}[4]{*}{n-Prec} & \multicolumn{2}{c|}{Post} & \multicolumn{2}{c|}{User} & \multicolumn{2}{c|}{Hashtag} \\
\cline{3-8}          &       & Base  & Ours  & Base  & Ours  & Base  & Ours \\
    \hline
    \multirow{4}[8]{*}{Shutdown} & 10-Prec & 1.000     & 1.000     & 0.900   & 1.000     & 1.000     & 1.000 \\
\cline{2-8}          & 50-Prec & 0.920  & 0.940  & 0.940  & 0.960  & 0.960  & 0.960 \\
\cline{2-8}          & 100-Prec & 0.870  & 0.930  & 0.850  & 0.920  & 0.910  & 0.930 \\
\cline{2-8}          & 200-Prec & 0.840  & 0.900   & 0.845 & 0.875 & 0.855 & 0.860 \\
    \hline
    \multirow{4}[8]{*}{Ebola} & 10-Prec & 1.000     & 1.000     & 0.800   & 1.000     & 1.000     & 1.000 \\
\cline{2-8}          & 50-Prec & 0.840  & 0.860  & 0.660  & 0.780  & 0.880  & 0.920 \\
\cline{2-8}          & 100-Prec & 0.770  & 0.840  & 0.660  & 0.730  & 0.870  & 0.880 \\
\cline{2-8}          & 200-Prec & 0.765 & 0.800   & 0.610  & 0.710  & 0.860  & 0.870 \\
    \hline
    \end{tabular}%
    }
    \caption{
\doc{Comparison of} the MRG computation method with the  baseline using top-n precision for posts, users, and hashtags.}
  \label{tab:Retrieval_Result_Evaluation}%
  \vspace{-2mm}
\end{table}%

There were too many posts, users\doc{,} or hashtags and we could not label all of them.
\kgsedit{Thus,} we did not report the recall in our evaluation.
In this evaluation, we used top n-precision (n-Prec) as the evaluation measure.
Top n-precision is the percentage of the \shimei{correctly retrieved} items among the top-n ranked items.
This measure is often used when the recall is hard to \shimei{calculate}~\cite{Chandramouli2007co}.
To fully compare the two algorithms, we calculated the top 10, 50, 100, \doc{and} 200-precision for posts, users, and hashtags, respectively.
\revision{We invited two PhD students who majored in data mining and are familiar with the datasets to evaluate the retrieval results.}
They labeled the \kgsedit{results individually} and resolved \doc{the} differences \doc{via} discussion.
The results are shown in Table~\ref{tab:Retrieval_Result_Evaluation}.
Overall, our algorithm performed better than the baseline on both datasets.
We \doc{inspected} the top 10 retrieved items \doc{with} both methods.
In general, \kgsedit{the retrieved items} \doc{were} quite accurate. \kgsedit{However,} the baseline \kgsedit{had} one mistake in the top 10 users selected from the shutdown dataset. It overestimated the importance of a user called @governmentclosd\doc{,} who posted \kgsedit{a significant number of} tweets with \kgsedit{a number of} hashtags.
\kgsedit{However,} this user did not have many followers and his/her tweets were seldom retweeted.
In contrast, our algorithm can avoid this mistake by taking a user's authority into consideration.
\kgsedit{The baseline algorithm also had similar mistakes in the} Ebola dataset.


\subsection{Incremental Ranking Update}
\revision{Since the incremental ranking update algorithm only calculate\shimei{s} the statistics of the changed random walks, it is more efficient than the full update.}
In this section, we conducted an experiment to \kgsedit{highlight} the effectiveness of our incremental ranking update algorithm.

\revision{First, we demonstrate \shimei{that} the incremental algorithm converges quickly. To this end, we} invited two analysts to use our system. One analyst \kgsedit{worked} on the Shutdown dataset \kgsedit{while} the other \kgsedit{worked} on the Ebola dataset.
\kgsedit{They updated the ranking incrementally based} on the initial retrieval \kgsedit{results.}
\doc{During} the \doc{update} process, when the analyst found that \doc{a} ranking score of an item \doc{was} \kgsedit{underestimated,} he \revision{increased} its ranking \shimei{score and} vice versa.
After each \kgsedit{update,} we re-calculated the top-200 precision for posts, users, and hashtags.
After \kgsedit{five updates,} we observed that results were nearly unchanged from the last update. \kgsedit{Hence, we allowed them to} stop the process.

The results after each update are listed in Table~\ref{tab:Incremental_Ranking_Update_Evaluation}.
It \kgsedit{shows that} the retrieval results \kgsedit{improved gradually} as they interactively modified the ranking scores.
This result verifies that our method can interactively refine the retrieval results by integrating \doc{analyst} feedback.

We can further observe that after some updates, the performance of more than one \doc{type} of \doc{item} change\shimei{d} \doc{as well}.
For example, after changing the first item in the Ebola dataset, the performance of the retrieved posts, users, and hashtags all increased. This result confirmed the effectiveness of the MRG model and the developed computation method.



\begin{table}[htbp]
  \centering
    \begin{tabular}{|c|c|c|c|c|c|c|}
    \hline
    \multirow{2}[4]{*}{Update} & \multicolumn{3}{c|}{Ebola} & \multicolumn{3}{c|}{Shutdown} \\
\cline{2-7}          & Post & User & Hashtag & Post & User & Hashtag \\
    \hline
    0     & 0.800   & 0.710  & 0.870  & 0.900   & 0.875 & 0.860 \\
    \hline
    1     & 0.815 & 0.715 & 0.875 & 0.910  & 0.875 & 0.865 \\
    \hline
    2     & 0.840  & 0.720  & 0.880  & 0.915 & 0.875 & 0.865 \\
    \hline
    3     & 0.855 & 0.720  & 0.885 & 0.925 & 0.885 & 0.875 \\
    \hline
    4     & 0.855 & 0.720  & 0.890  & 0.925 & 0.885 & 0.880 \\
    \hline
    5     & 0.855 & 0.720  & 0.895 & 0.925 & 0.885 & 0.885 \\
    \hline
    \end{tabular}%
    \caption{Evaluation of our incremental ranking update algorithm.
    The changes are evaluated by top-200 precision.
    The row with 0 \doc{updates} contains the initial \kgsedit{results}.}
  \label{tab:Incremental_Ranking_Update_Evaluation}%
  \vspace{-2mm}
\end{table}%

\revision{\shimei{Second, since} the incremental update algorithm can fully \doc{update} the statistics of the changed random walks, the incremental update achieves the same \shimei{ranking} result as the full update \shimei{algorithm}.}

\section{Application}\label{sec:application}

\kgsedit{In order to} evaluate the usefulness of MutualRanker, we \shimei{performed} two case studies on \shimei{the same} Twitter \shimei{datasets} described in Sec.~\ref{sec:evaluation}.
\shimei{Due to the page \doc{limit},} we focus our report \shimei{on the} shutdown dataset.
Interested readers \shimei{may refer} to the attached video for the \shimei{study on} the Ebola dataset.
\revision{\shimei{Moreover,} MutualRanker allows users to filter out irrelevant items based on \shimei{their} knowledge.
For example, in the government shutdown case study, \shimei{users can remove} irrelevant hashtags such as ``\#retweet,'' ``\#rt,'' ``\#path,'' and ``\#road'' from the initial query.}

\revision{The procedure of the case studies was loosely structured into three phases.
First, we pre-interviewed two experts, one researcher in sociology (S) and one researcher in media and communication\doc{s} (C), to understand their respective interests in the datasets.
We designed a number of exploration tasks.
In the second phase, we collaborated with the experts to finish the designed tasks.
During this phase, we asked questions to discuss with the experts the usefulness of our tool for each task.
Finally, the experts were invited to another discussion session to provide overall feedback on how our tool could help them with real-world tasks.\looseness=-1}

\subsection{Case Study: Government Shutdown}


\revision{In \shimei{this study}, we worked \doc{with expert} \shimei{S} to: 1) evaluate how uncertainty analysis can be utilized to identify key hashtags and users with a satisfactory confidence level;
2) leverage our system to iteratively reduce the uncertainty levels;
3) extract relevant hashtags/users/tweets \shimei{related to} the government shutdown.}

\noindent \textbf{\normalsize Overview}. The expert quickly found interesting results after examining the hashtag overview (Fig.~\ref{fig:shutdown_1}(a)) generated by our system.
She identified seven prominent topics described by a set of hashtags: general discussions about the shutdown and Obamacare
(Fig.~\ref{fig:shutdown_1}A), political \kgsedit{discourse} on twitter (Fig.~\ref{fig:shutdown_1}B),
discussion on ending \kgsedit{the} shutdown (Fig.~\ref{fig:shutdown_1}C),
\kgsedit{the} influence of \kgsedit{the} shutdown \kgsedit{on} people's \kgsedit{lives} (Fig.~\ref{fig:shutdown_1}D)
reporting \kgsedit{the} government shutdown on news media (Fig.~\ref{fig:shutdown_1}E), \kgsedit{debt-related} discussion (Fig.~\ref{fig:shutdown_1}F), and critics \kgsedit{of the} shutdown (Fig.~\ref{fig:shutdown_1}G)\revision{.}



\begin{figure}[t]
\centering
\vspace{-2mm}
\includegraphics[width=0.45\textwidth]{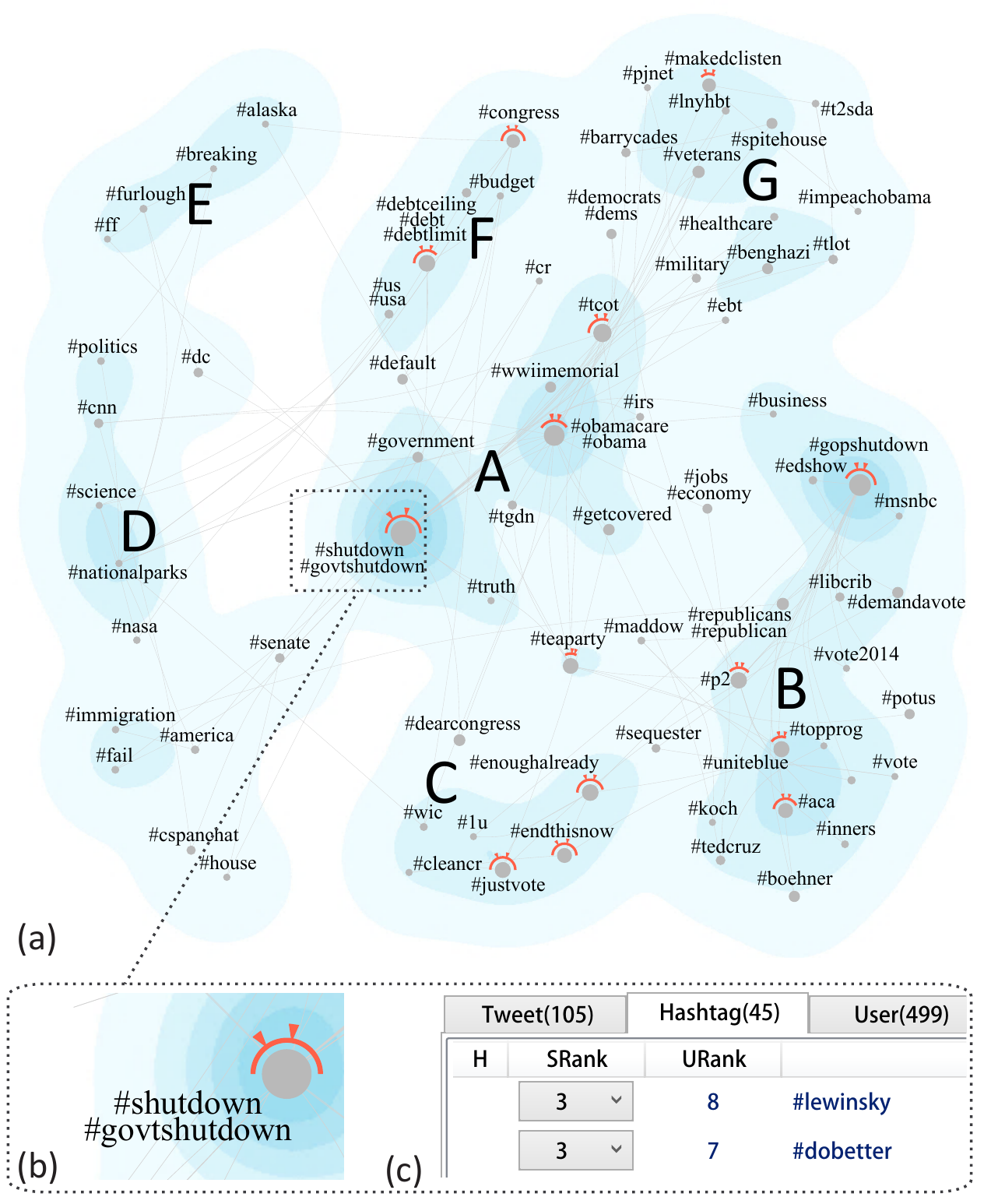}
\vspace{-1mm}
\caption{The overview of the government shutdown dataset.}
\vspace{-5mm}
\label{fig:shutdown_1}
\end{figure}

\noindent \textbf{\normalsize Uncertainty analysis}:
\revision{The ``\#shutdown'' cluster (Fig.~\ref{fig:shutdown_1}(b)) attracted the expert's attention because it contains items with higher uncertainty.}
The expert examined the detailed hashtags and tweets in the cluster.
She found that in addition \kgsedit{to} common hashtags \kgsedit{such as} \#govtshutdown, \#obamashutdown, \doc{and} \#shutdowngop, \shimei{a number of diverse hashtags} were also created. Such hashtags included those that criticized \shimei{the shutdown}, e.g., \#shutdownharry; local news posts, e.g., \#hounews, and public campaigns, e.g., \kgsedit{\#dontcutkids}.
She wanted to examine the most uncertain ones, so she sorted the hashtags by the uncertainty level.
Interestingly, \#lewinsky was ranked \shimei{as} the \doc{most} uncertain hashtag (Fig.~\ref{fig:shutdown_1}(c)).
\revision{The analyst searched the related tweets and found that data tagged with \#lewinsky concerned the shutdown of the Clinton government in 1995.}
The expert decided to lower the ranking score of the hashtag.
During the process, she commented that the uncertainty glyph and \doc{item-filtering} feature were useful, \doc{helping} her filter out irrelevant items by lowering their ranking scores.





\noindent \textbf{\normalsize Uncertainty propagation}: Next, the expert examined how the uncertainty of the ``\#shutdown'' cluster \revision{would} influence neighboring \kgsedit{clusters}.
She clicked the ``propagation'' button and the corresponding uncertainty propagation \kgsedit{was} displayed (the orange flow in Fig.~\ref{fig:shutdownoverview}).
She also selected the uncertainty propagation of the ``\#democrats'' cluster (the blue flow in Fig.~\ref{fig:shutdownoverview}) and ``\#republicans'' cluster (the \kgsedit{green} flow in Fig.~\ref{fig:shutdownoverview}), which \doc{were} closely related to the ``\#shutdown'' cluster.
As shown in Fig.~\ref{fig:shutdownoverview}(b), cluster ``\#nationalparks'' shared the uncertainty propagated \shimei{from} the three \shimei{clusters}.
\kgsedit{Given that} the closing of the national parks was a result of the government shutdown and \kgsedit{stimulated discussion} on \kgsedit{Twitter}, the expert increased the ranking score of \#nationalparks (from 4 to 6).
In our system, the ranking score is from 1 to \kgsedit{10,} with 10 being the highest score.\looseness=-1

After the adjustment, she noticed the scores of another two hashtag clusters were automatically increased: ``\#spitehouse'' and ``\#teaparty.''
In the first cluster, the ranking scores of hashtags \kgsedit{such as} ``\#spitehouse'' and ``\#demshutdown'' \kgsedit{increased.}
In the second \kgsedit{cluster}, the ranking scores of hashtags \kgsedit{such as} ``\#teaparty'' and ``\#defundgop'' increased, \kgsedit{as well}.
The expert commented, ``It is helpful that the hidden relationships between hashtags are leveraged to propagate the ranking change.
I can find more partisan messages around the topic and the public responses in this way.''
She then found related tweets in the \#spitehouse group and \#teaparty group.
For example, ``@RepBradWenstrup @sarahlance \#shutdown \#Nationalpark Here's what my tea-party-backed \#Republican did to my vacation.''


On the contrary, the ranking score of cluster ``\#ebt'' decreased, which is caused by \kgsedit{the} ranking score decrease of hashtags ``\#ebt'' and ``\#\kgsedit{obamzombies}.''
The expert then examined the relevant tweets to probe the reason.
The \kgsedit{EBT system} was crashed \kgsedit{at that time} and \kgsedit{many} people wondered whether the crash was caused by the government shutdown:
``Ahh... \#ebt not working cause if a \#governmentshutdown? How sad you can't spend money taken from me against my will that I worked for\ldots''
\kgsedit{Then,} the crash was \kgsedit{explained to be} a result of a computer failure (``According to NBC, \#ebt is down because of a technical issue, NOT \#governmentshutdown'').
\revision{Thus, the expert believed ``\#ebt'' was \shimei{irrelevant and} appreciated this automatic change.}


\noindent \textbf{\normalsize Switching between different data views.} In addition to hashtags, the expert wanted to examine the users who participated in different discussion groups.
For example, she wanted to \kgsedit{identify} the most active users in the ``\#shutdown'' cluster, so she \kgsedit{overlaid} the user labels around the hashtag labels (Fig.~\ref{fig:usercontext}(a)).
The expert then switched to the user view to explore \kgsedit{additional} user information (Fig.~\ref{fig:usercontext}(b) and Fig.~\ref{fig:usercontext}(c)).
She immediately identified the leading users in Fig.~\ref{fig:usercontext}(b) and Fig.~\ref{fig:usercontext}(c).
She described them with two categories: 1) \kgsedit{key} government official accounts, including ``@barackobama,'' ``@whitehouse'' (Fig.~\ref{fig:usercontext}(b)); and 2) news agencies/public media \kgsedit{such as} ``@nytimes,'' ``@guardian,'' and ``@bloombergnews'' (Fig.~\ref{fig:usercontext}(c)).
\kgsedit{Considering that} partisan leaders were \kgsedit{of} major interest to her, \shimei{she} first observed the ranking scores of \kgsedit{select} politicians, e.g., @speakerboehner (Rank 8), @whiphoyer (Rank 8), @nancypelosi (Rank 7)\doc{,} etc.
She believed that the importance of these user accounts \doc{was underestimated} \kgsedit{because} the influence and activeness of politicians on twitter \doc{are} usually much lower than \kgsedit{that} in real life.
She changed the rankings of the partisan leaders, ``@speakerboehner,'' ``@whiphoyer,'' and ``@nancypelosi,'' to 10, which is the highest.
Fig.~\ref{fig:usercontext}(d) shows the difference after this refinement.

\begin{figure}[t]
\centering
\vspace{-2mm}
\includegraphics[width=0.45\textwidth]{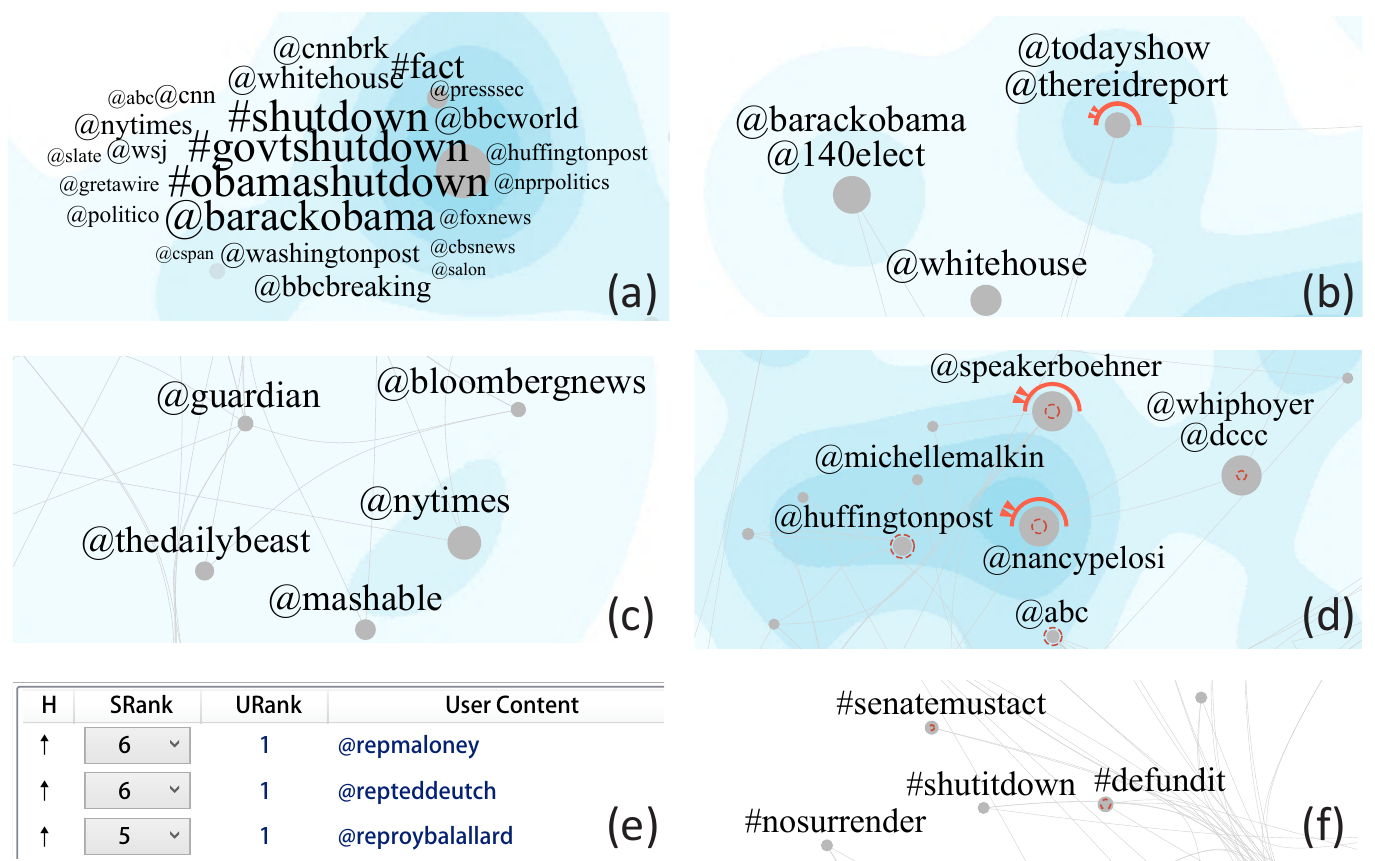}
\vspace{-1mm}
\caption{
Switching between different \revision{data graphs}: (a) overlay labels around the focus cluster; (b) key government official accounts in the \revision{user graph}; (c) news agencies/public media in the \revision{user graph}; (d) the \revision{user graph} after \doc{changing}; (e) new ranking scores of a few politicians; (f) switch back to the \revision{hashtag graph}.
}
\vspace{-5mm}
\label{fig:usercontext}
\end{figure}

After the change, the user clusters were regenerated and the uncertainly levels of some nodes were largely reduced.
Notably, ``@whiphoyer'' became an important cluster with the scores of several users in the cluster automatically increased (Fig.~\ref{fig:usercontext}(e)). For example, ``@repmaloney,'' from 5 to 6 and ``@repteddeutch,'' from 5 to 6.
``These are members of Congress. The change of their ranking scores is natural here.'' The expert commented\doc{,} ``This is cool.
[...]
If I want to change the ranking score of one user, others just automatically follow.
This could help me find \doc{the} important users whose names I am not familiar with or who are not active on Twitter.''

The \kgsedit{expert} then switched back to the \revision{hashtag graph} to check the influence of the change on this graph.
She \kgsedit{found} a new hashtag cluster, ``\#senatemustact.''
She then zoomed into this cluster.
As shown in Fig.~\ref{fig:usercontext}(f), \kgsedit{the hashtag primarily expresses criticism of} the government\kgsedit{, blaming either the Democrats or Republicans} (``@PeteSessions \#DefundObamacare \#shutdown \#MakeDCListen \#senatemustact Stand for the American People!'').


\subsection{User Feedback}

\revision{To evaluate the usefulness of our system, we conducted a semi-structured interview with the two experts.
\doc{They used} MutualRanker in the case \doc{study} for 2 hours, so they \doc{were} familiar with its basic functions.}
Overall, MutualRanker was well received by them.

The experts appreciated MutualRanker as a research tool to help them collect relevant posts, users, and hashtags quickly and conveniently.
Expert C believed that MutualRanker is very useful for coding in media and communication\doc{s}.
According to him, coding is the most \kgsedit{labor-intensive} work in his \kgsedit{field}. Extensive training and careful attention have always been required to produce reliable data.
He commented, ``A toolkit like MutualRanker is urgently needed in my daily work to reduce coding complexity and costs.
Especially when there are not enough samples, the linkage between items [in this system] will provide more information  to make decision. [...] This system also provides an opportunity to supervise the data retrieval process.''

\doc{Both} the experts were impressed by the uncertainty illustration and its propagation function.
For example, Expert C said, ``Uncertainty propagation is an awesome feature, I can use it to find some unexpected data and increase the coverage of coding.''

The experts agreed that smoothly switching between different data graphs \kgsedit{helped} them find relevant data more quickly.
Expert S commented, ``This switching function enables me to easily transition between \revision{the hashtag graph and the user graph.}
When I modify one ranking score in one graph, I \doc{cannot} only verify the result in this graph, but \doc{also verify} it in another graph. ''

The experts also suggested several improvements.
The target \kgsedit{audience} of MutualRanker is experts with domain knowledge.
The experts believed that average users \shimei{can also benefit from it}.
They \kgsedit{suggested that} more intuitive visual design \kgsedit{be used}.
Expert C said, ``The uncertainty glyph can be simplified for a general user. For example, maybe the glyph \doc{does} not need to encode the uncertainty distribution, just simply show that this ranking score is uncertainty.''
They also expressed the need to retrieve streaming data.


\section{Discussion and Future Work}\label{sec:conclustion}
This paper presents a visual analytics system, MutualRanker, to help analysts interactively \revision{retrieve} data of interest from microblogs.
We extend the \kgsedit{MRG} model to extract a multifaceted retrieval result \kgsedit{that includes} the mutual reinforcement ranking results, the uncertainty of each rank, and the uncertainty propagation among different graph nodes.
The model is tightly integrated with a composite visualization to \kgsedit{assist} analysts \kgsedit{in retrieving} salient posts, users, and hashtags \kgsedit{effectively}, in an uncertainty-aware environment.

In the future, we plan to improve system performance by \shimei{implementing a parallel} Monte Carlo sampling method.
Another exciting avenue for future work is to retrieve streaming data in microblogs, which \kgsedit{can be} very useful in emergency management and threat analysis.
We believe the system can also benefit average \kgsedit{users} interested in collecting microblog data.
\kgsedit{In the future, we} will \kgsedit{also} invite more users to \kgsedit{try} our system and conduct a formal user study.
Accordingly, we \kgsedit{will} improve MutualRanker based on the collected feedback.
}

\acknowledgments{
We would like to thank X. Wang and J. Yin, J. Gong, and Dr. W. Cui for helpful discussions on the visualization design, Dr. J. Zhang and Dr. Y. Song for constructive suggestions on similarity measures, as well as Dr. W. Peng and Dr. J. Su for providing domain expertise.\looseness=-1
}
\small
\bibliographystyle{abbrv}
\bibliography{reference}



\end{document}